\documentclass[10pt,twocolumn,showpacs,preprintnumbers,amsmath,amssymb,amsfonts,superscriptaddress,floatfix,prl,aps]{revtex4-1}
\usepackage{CJK}

\usepackage[paperwidth=210mm,paperheight=297mm,centering,hmargin=2cm,tmargin=1.6cm,bmargin=3.cm]{geometry}

\usepackage[dvipsnames,table,xcdraw]{xcolor}

\usepackage{hyperref}
\hypersetup{
    colorlinks=true,
    linkcolor=Blue,
    filecolor=Blue,      
    urlcolor=Blue,
    citecolor=blue,
}

\usepackage{enumitem}
\usepackage{booktabs}
\usepackage{subfigure}
\usepackage{graphicx}
\usepackage{amsfonts}
\usepackage{amssymb}
\usepackage{amsmath}
\usepackage{overpic}
\usepackage{siunitx}

\usepackage{braket}
\usepackage{xcolor}
\usepackage{cancel}
\usepackage{fancyhdr}
\usepackage[utf8]{inputenc}
\usepackage{lipsum}
\usepackage[T1]{fontenc}
\usepackage[newcommands]{ragged2e}
\usepackage[font=small,labelfont=bf,justification=Justified]{caption}
\usepackage{xcolor}
\usepackage[english]{babel}

\newcommand{\beq}{\begin{equation}}
\newcommand{\eeq}{\end{equation}}
\newcommand{\bc}{\begin{center}}
\newcommand{\ec}{\end{center}}
\newcommand{\bee}{\begin{eqnarray}}
\newcommand{\eee}{\end{eqnarray}}

\newcommand{\om}{\omega}
\newcommand{\s}{\sigma}
\newcommand{\bb}{\textbf}

\makeatletter
\makeatletter \renewcommand{\fnum@figure}{{\bf{\figurename~\thefigure}}}

\makeatother

\begin{document}

\title{ 
Tuning macroscopic phase frustration in multiorbital superconductors}

\author{I. Maccari} 
\email{imaccari@phys.ethz.ch}
\affiliation{Institute for Theoretical Physics, ETH Zurich, CH-8093 Zurich, Switzerland}
\author{A. Ramires}
\email{aline.ramires@tuwien.ac.at}
\affiliation{Institute of Solid State Physics, TU Wien,
1040 Wien, Austria}

\date{\today}

\begin{abstract}
{Time-reversal symmetry-breaking (TRSB) superconductivity has been reported in a growing number of materials. 
In some cases, TRSB arises naturally from chiral superconductivity, but in many low-symmetry systems this explanation is not viable. 
In these latter cases, TRSB is often attributed to phase frustration among multiple superconducting gaps on different Fermi surfaces. 
Yet, the microscopic conditions enabling such frustration remain poorly understood. 
Here, inspired by the TRSB reported in the superconducting state of iron-based materials, we demonstrate that a minimal two-orbital model can support a TRSB superconducting state via phase frustration. 
We identify the key microscopic parameters that stabilize TRSB in d-electron systems with orthorhombic symmetry and provide a framework to systematically enlarge the region of parameter space within which TRSB is expected in materials with other electronic content and crystalline symmetries. 
Our results offer a simple and experimentally relevant route to understand and control TRSB in multiorbital superconductors.}
\end{abstract}
\maketitle

\emph{Introduction} -- 
Time-reversal symmetry-breaking (TRSB) superconductivity has been reported in many different families of materials \cite{ghosh_recent_2020}. 
In ruthenates \cite{luke_time-reversal_1998, xia_highresolution_2006} and Kagome systems \cite{mielke_time-reversal_2022, guguchia_tunable_2023,deng_evidence_2024, graham_pressure_2025}, TRSB can be ascribed to chiral superconductivity, as the underlying crystal structures have point group symmetries that support two-dimensional irreducible representations \cite{sigrist_phenomenological_1991, kallin_chiral_2016, ramires_symmetry_2022}. 
In contrast, UTe$_2$ \cite{hayes_multicomponent_2021,wei_interplay_2022}, LiNiGa$_2$ \cite{hillier_nonunitary_2012, weng_twogap_2016, ghosh_quantitative_2020}, and LiNiC$_2$ \cite{hillier_evidence_2009, quintanilla_relativistic_2010} were also reported to break time-reversal symmetry at the superconducting critical temperature, but their orthorhombic point groups do not support multi-dimensional irreducible representations giving rise to symmetry-related multi-component superconductivity.
The resolution to the conundrum of TRSB superconductivity in these latter cases might rely on other flavors of TRSB superconductivity, based on the notion of nonunitarity and phase frustration. 

{Nonunitary superconductivity arises when the superconducting wave function acquires an internal polarization—such as spin, orbital, or sublattice— which may lead to TRSB.
Such states necessarily involve multiple superconducting components associated with different types of pairs formed across distinct internal degrees of freedom.}
In its simplest form, nonunitary superconductivity occurs in the spin-triplet sector of single-band systems \cite{sigrist_phenomenological_1991, hillier_nonunitary_2012}. 

{A complementary perspective on TRSB builds on the concept of phase frustration, in which multiple scalar order parameters—typically associated with distinct bands—are coupled through Josephson-like interband interactions that favor incompatible relative phases.
The resulting complex phase differences lead to spontaneous TRSB \cite{tanaka_chiral_2010, stanev_three-band_2010, hu_stability_2012, maiti_s+is_2013, garaud_domain_2014, tanaka_multicomponent_2015}. }
Recently, the proposal of a generalized notion of nonunitary superconductivity has shown that these two perspectives can be encapsulated within a single framework \cite{ramires_nonunitary_2022}. However, {in the materials listed above}, the microscopic conditions that promote TRSB  through these latter mechanisms remain poorly understood.

The understanding of the origin of TRSB superconductivity through phase frustration can be deepened by exploring its dependence on certain tuning parameters. 
In this context, materials in the family of iron-based superconductors are relevant as these have also been reported to break time-reversal symmetry \cite{grinenko_superconductivity_2017-1, grinenko_superconductivity_2020, grinenko_state_2021, matsuura_two_2023, roppongi_topology_2025}, with some manifesting TRSB only within a relatively narrow doping range. 
Particularly striking is the example of Ba$_{1-x}$K$_x$Fe$_2$As$_2$: by varying the hole doping $x$, the system is reported to break time-reversal symmetry only for $0.7 \lesssim x \lesssim 0.85$, near a Lifshitz transition
\cite{grinenko_superconductivity_2017-1, grinenko_superconductivity_2020}.
Intriguingly, around $x \simeq0.77$, the TRSB signal persists even above the superconducting state, consistent with the possible emergence of a quartic metal phase formed by the condensation of four electrons into a fermion quadrupling superfluid \cite{grinenko_state_2021, shipulin_calorimetric_2023, halcrow_probing_2024}.

The origin of TRSB in this system is generally attributed to phase frustration arising from repulsive Josephson interactions among superconducting order parameters that develop in three different bands \cite{stanev_three-band_2010, carlstrom_length_2011, hu_stability_2012, maiti_s+is_2013, boker_s+is_2017, grinenko_superconductivity_2020}. 
Microscopically, these repulsive interactions are thought to originate from pair-hopping processes \cite{stanev_three-band_2010, maiti_s+is_2013}, which favor opposite signs of the gaps on three different Fermi pockets. 
Additionally, the shift of the TRSB superconducting state slightly away from the Lifshitz transition, at which the electron pockets vanish, has been ascribed to a renormalization of the chemical potential \cite{boker_s+is_2017}. 

{Despite significant insights gained from studies of iron-based superconductors, the role of the normal-state Hamiltonian in stabilizing TRSB superconductivity remains underexplored. 
In particular, the influence of microscopic parameters encoded in the orbital basis, and how these generalize to a broader class of materials, is still poorly understood. 
In this Letter, we develop a microscopic framework to clarify how a minimal two-orbital model can support TRSB superconductivity, and we identify the key conditions—band filling, spin–orbit coupling (SOC), and orthorhombic distortions—that stabilize this state in
d-electron systems with low crystalline symmetry.} 

\emph{Microscopic model} --
To explore how frustration can be tuned to stabilize a TRSB superconducting order parameter, we derive the Landau free energy from a microscopic two-orbital model. 
The most general normal metal Hamiltonian describing a system with two orbitals and two spins can be written as
\beq
\hat{H}_0(\bb{k})= \sum_{(a,b)} h_{ab}(\bb{k}) \hat{\tau}_a \otimes \hat{\s}_b,
\label{H_orbspin}
\eeq
where $\hat{\tau}_a$  and $\hat{\s}_b$ are $2 \times 2$ identity matrices, $\{a,b\}=0$, or Pauli matrices, $\{a,b\} = \{1,2,3\}$, serving as basis for the orbital and spin degrees of freedom, respectively.
We consider an effective zero-momentum pairing interaction of the form
\beq
\begin{split}
H_{int} = &\sum_{\bb{k}, \bb{k}', \{\alpha\}, \{\s\}} V_{\{\alpha, \s\}}(\bb{k}, \bb{k}') \times\\
&   c^{\dagger}_{\alpha_1, \s_1, \bb{k}}
c^{\dagger}_{\alpha_2, \s_2, -{\bb{k}}}
c_{\alpha_3, \s_3, \bb{k}'}
c_{\alpha_4, \s_4,-{\bb{k}'}},
\end{split}
 \eeq
where $c^\dagger_{\alpha, \s, \bb{k}}$ ($c_{\alpha, \s, \bb{k}}$) creates (annihilates) an electron with orbital index $\alpha$, spin $\s$, and momentum $\bb{k}$, and $\{\alpha,\sigma\}$ denotes all summed indices.

By performing the standard Hubbard–Stratonovich transformation and integrating out the fermionic degrees of freedom, the quadratic contribution to the Landau free energy can be written as:
\begin{eqnarray}
&&\mathcal{F}^{2nd} =   -\frac{1}{\beta} \sum_{\bb{k},\om_m}   Tr\left[\hat{{G}}_0 (\bb{k},\omega_n) \hat{\Delta}(\bb{k})\hat{\bar{G}}_0 (\bb{k},\omega_n)\hat{\Delta}^{\dagger}(\bb{k})\right]  \nonumber \\ &&-\sum_{\bb{k}, \bb{k}', \{\alpha,\s\}}  V_{\{\alpha, \s\}}(\bb{k},\bb{k}')\Delta^{\s_1 \s_2}_{\alpha_1 \alpha_2}(\bb{k}) \Delta^{\s_3 \s_4 *}_{\alpha_3 \alpha_4}(\bb{k}'),
\label{f_GL}
\end{eqnarray}
where $\Delta^{\s_1 \s_2}_{\alpha_1 \alpha_2}(\bb{k})$ denotes the superconducting gap between a pair of electrons in states $\{\alpha_1, \s_1, \bb{k}\}$ and $\{\alpha_2, \s_2, -\bb{k}\}$, $\beta=1/(k_\mathrm{B}T)$ is the inverse temperature, $\omega_m=\pi k_B T(2m+1)$ is the fermionic Matsubara frequency, and $k_B$ is the Boltzmann constant. The non-interacting Green's function matrices read $\hat{G}_0(\bb{k},\omega_m)=[i\om_m\hat{\mathbb{I}} - \hat{H}_0(\bb{k})]^{-1}$ and $\hat{\bar{G}}_0(\bb{k},\omega_m)=[i\om_m\hat{\mathbb{I}} + \hat{H}^*_0(-\bb{k})]^{-1}$. 
In the presence of both time-reversal and inversion symmetry, only a subset of six pairs of indices $(a,b)$ are symmetry allowed in $\hat{H}_0(\bb{k})$, guaranteeing two doubly degenerate eigenvalues, $
E_{1,2}(\bb{k}) = h_{00}(\bb{k}) \pm |\bb{h}(\bb{k})|$, where $\bb{h}(\bb{k})$ is the five-dimensional vector formed by the terms ${h_{ab}}(\bb{k})$ with $(a,b)\neq(0,0)$. 
Under these considerations, the non-interacting Green's function matrices can be conveniently written as 
\beq
\begin{split}
\hat{G}_0(\bb{k}, \om_m) = P_1 (\bb{k}) g_1(\bb{k}, \om_m)+ P_2(\bb{k}) g_2(\bb{k}, \om_m),
\end{split}
\label{G0}
\eeq
 where $g_n(\bb{k}, \om_m)= [i\om_m -E_n(\bb{k})]^{-1}$ and $P_{1,2}(\bb{k}) = [1 \pm \tilde{H}(\bb{k})]/2$ are the projector operators into the respective bands. Here  $\tilde{H}(\bb{k})=  \delta \hat{H} (\bb{k})/ {|\bb{h}(\bb{k})|}$ and $  \delta \hat{H} (\bb{k}) = \hat{H}_0( \bb{k}) - h_{00}(\bb{k})\hat{\tau}_0\otimes \hat{\sigma}_0$.
Finally, for a two-orbital system, the superconducting order parameter can be expressed in matrix form as
\beq
\hat{\Delta}(\bb{k}) = \sum_{(a,b)} d_{a b}(\bb{k}) \hat{\tau}_a \otimes \hat{\sigma}_b (i \hat{\sigma}_2);
\label{Delta_dab}
\eeq
where, as in Eq.~\eqref{H_orbspin}, $\hat{\tau}_a$ and $\hat{\sigma}_b$ parametrize the orbital and spin degrees of freedom, respectively.
The gauge-invariant product
of the gap matrix, 
\beq
\hat{\Delta}(\bb{k}) \hat{\Delta}^{\dagger}(\bb{k}) = |\Delta_U(\bb{k})|^2 \hat{\tau}_0 \otimes \hat{\s}_0 + q^{(ab)}_{NU} (\bb{k})  \hat{\tau}_a \otimes \hat{\s}_b ,
\label{gaugeinv}
\eeq
can be separated in unitary and nonunitary parts.
{
The unitary part is given by
\beq
 |\Delta_U(\bb{k})|^2 = \sum_{(a,b)}|d_{ab} (\bb{k})|^2; 
\label{Deltak_U}
\eeq
while the nonunitary part can generally be decomposed as
\beq
   q^{(ab)}_{NU} (\bb{k}) = q^{(ab)}_{TRE,NU}(\bb{k})+ {q}^{(ab)}_{TRO, NU}(\bb{k}),
    \label{Deltak_NU}
\eeq
where the first and second term correspond, respectively, to the time-reversal-even (TRE) and time-reversal-odd (TRO) contributions~\cite{brydon_bogoliubov_2018}.
For later convenience, we further separate the contributions involving the $d_{00}$ orbital component as $q^{(ab)}_{TRE,NU}(\bb{k})= q^{(ab)}_{0,NU} + q^{(ab)}_{1,NU} $, where $q^{(ab)}_{0,NU}$ always takes the form
\beq
    q^{(ab)}_{0,NU}(\bb{k})= d_{00}(\bb{k}) d^*_{ab}(\bb{k}) + c.c..
    \label{Deltak_NU0}
\eeq
Assuming inversion symmetry, we focus on the superconducting order parameters in the even-parity sector, where ${q}^{(ab)}_{1,NU}(\bb{k})=0$.
The explicit form of ${q}^{(ab)}_{TRO, NU} (\bb{k})$ for the case of two orbitals with equal parity is~\cite{ramires_nonunitary_2022}
\begin{eqnarray}
    &&{q}^{(ab)}_{TRO, NU} (\bb{k})\tau_a \otimes \sigma_b = i(d_{30}d^*_{10} -c.c.) \tau_2 \otimes \sigma_0 +\nonumber \\
    && i\sum_{b\neq0} \left[(d_{10}d^*_{2b} -c.c.) \tau_3 + (d_{2b}d^*_{30} -c.c.) \tau_1  \right] \otimes \sigma_b +\nonumber \\
    &&  i \sum_{a\neq b\neq 0} (d_{2a}d°*_{2b} - c.c.)\varepsilon_{abc}\tau_0 \otimes \sigma_c.
    \label{Deltak_NUTRO}
\end{eqnarray}
In Eq.\eqref{Deltak_NUTRO}, the three lines correspond, respectively, to an orbital, spin-orbit, and spin polarization of the superconducting wave function.
}

By means of Eqs.\eqref{G0} and \eqref{Delta_dab}, we can rewrite the first term on the right side of Eq.\eqref{f_GL} as
\begin{widetext}
\begin{equation}
    \begin{split}
     - k_B T \sum_{\bb{k},\omega_m} Tr&\left[ \hat{{G}}_0 \hat{\Delta}\hat{\bar{G}}_0 \hat{\Delta}^{\dagger} \right]=  -\frac{\mathcal{L}_1(T)}{8} \left[ N_1  \Big\langle \left(T_1 + T_4\right) + (T_2 + T_3) \Big\rangle_{FS_1} +  N_2  \Big\langle (T_1 + T_4) - (T_2 + T_3) \Big\rangle_{FS_2} \right]+ \\ & + \mathcal{L}_2(T) \left[N_1\left\langle \frac{(T_1 -T_4)+ (T_2 + T_3)}{|\bb{h}|^2}\right\rangle_{FS_1} + N_2 \left\langle \frac{(T_1 -T_4) -  (T_2 + T_3)}{|\bb{h}|^2} \right\rangle_{FS_2}\right],
    \end{split}
\label{2ndorderGL_T}
\end{equation}
\end{widetext}
where we omitted the explicit $\bb{k}$ dependencies for conciseness. $N_1, N_2$ are the density of states at the Fermi level of the two bands, while $\langle \dots \rangle_{FSi} $ indicates the angular average on the Fermi surface of the band $i=1,2$. $\mathcal{L}_1(T)=\ln \left( \frac{4e^\gamma}{\pi} \frac{\omega_c}{2k_B T} \right)$, and $\mathcal{L}_2(T)={\omega_c^2}\tanh{(\frac{\omega_c}{2k_B T})}$ were obtained by performing the sum over momenta and Matsubara frequencies, i.e. $T\sum_{\omega_m,\bb{k}} g_{n_1}\bar{g}_{n_2}$, and assuming that the separation between the two bands is much larger than the energy cutoff, i.e. $|\bb{h}(\bb{k})| \gg \omega_c$. The traces $T_i$ in Eq.\eqref{2ndorderGL_T} can be written in terms of the unitary, Eq.\eqref{Deltak_U}, {and nonunitary, Eq.  \eqref{Deltak_NU0}}, part of the superconducting gap:
\begin{eqnarray}
\label{t1}
    T_1 &&= Tr \left\{\hat{\Delta} \hat{\Delta}^{\dagger} \right\} = 4|\Delta_U (\bb{k})|^2;\\
     \label{t2}
     T_2&&=Tr \left\{ \tilde{H}\hat{\Delta} \hat{\Delta}^{\dagger}  \right\}= - 4\sum_{(a,b)'}q^{(ab)}_{0,NU}\tilde{h}_{ab}(\bb{k});\\
     \label{t3}
    T_{3} &&= Tr \left\{  \tilde{\bar{H}} \hat{\Delta}^{\dagger}\hat{\Delta}\right\} = - 4\sum_{(a,b)'}q^{(ab) \dagger}_{0,NU}\tilde{h}_{ab}(\bb{k});\\
     \label{t4}
    T_4&&=  Tr \left\{ \tilde{H} \hat{\Delta}\tilde{\bar{H}} \hat{\Delta}^{\dagger} \right\} = 8 |d_{00}|^2 - 4|\Delta_U|^2 + \nonumber \\ &&\,
    8 \sum_{(a,b)'} \sum_{(c,d)'} \tilde{h}_{ab}(\bb{k})\tilde{h}_{cd}(\bb{k}) d_{ab}(\bb{k})  d^*_{cd}(\bb{k}) .
\end{eqnarray}
Here and in what follows, 
$\tilde{h}_{ab}(\bb{k}) = {h_{ab}(\bb{k})}/{|\bb{h}(\bb{k})|}$. 
{The expressions for the traces in Eq.~\eqref{t1}-\eqref{t4} hold generically for a two-orbital system with inversion and time-reversal symmetry.} 
From the explicit form of the traces, we can already draw the following general conclusions:
    (i) the trace $T_1$ only depends on the unitary part of the superconducting gap;
    (ii) the traces $T_2$ and $T_3$ are non-zero only if the superconducting gap has a nonunitary component, i.e. $q^{(ab)}_{0,NU}\neq0$;
    (iii) since $q^{(ab)}_{0,NU}0$ is real, $T_2= T_3$;
    (iv) the relative phase differences between the $d_{ab}(\bb{k})$ components arise from $T_{2,3}$ and $T_4$. The former traces constrain the phase only between $d_{00}(\bb{k})$ and $d_{ab}(\bb{k})$ with $(a,b)\neq(0,0)$.

With the explicit form of Eq.~\eqref{2ndorderGL_T}, we can capture {part of the effective Josephson couplings between different superconducting components
$ \propto J_{(ab), (dc)} (d_{ab} d^*_{dc} + c.c.)$ contributing to the Landau free energy [in addition to the contribution stemming from the interaction vertex in the second line of Eq. \eqref{f_GL}].}
In the following, we examine a specific case that provides a theoretical framework for the emergence of a nonunitary superconducting gap spontaneously breaking time-reversal symmetry in orthorhombic systems.

\begin{figure}[t!]
    \centering
    \includegraphics[width=0.8\linewidth]{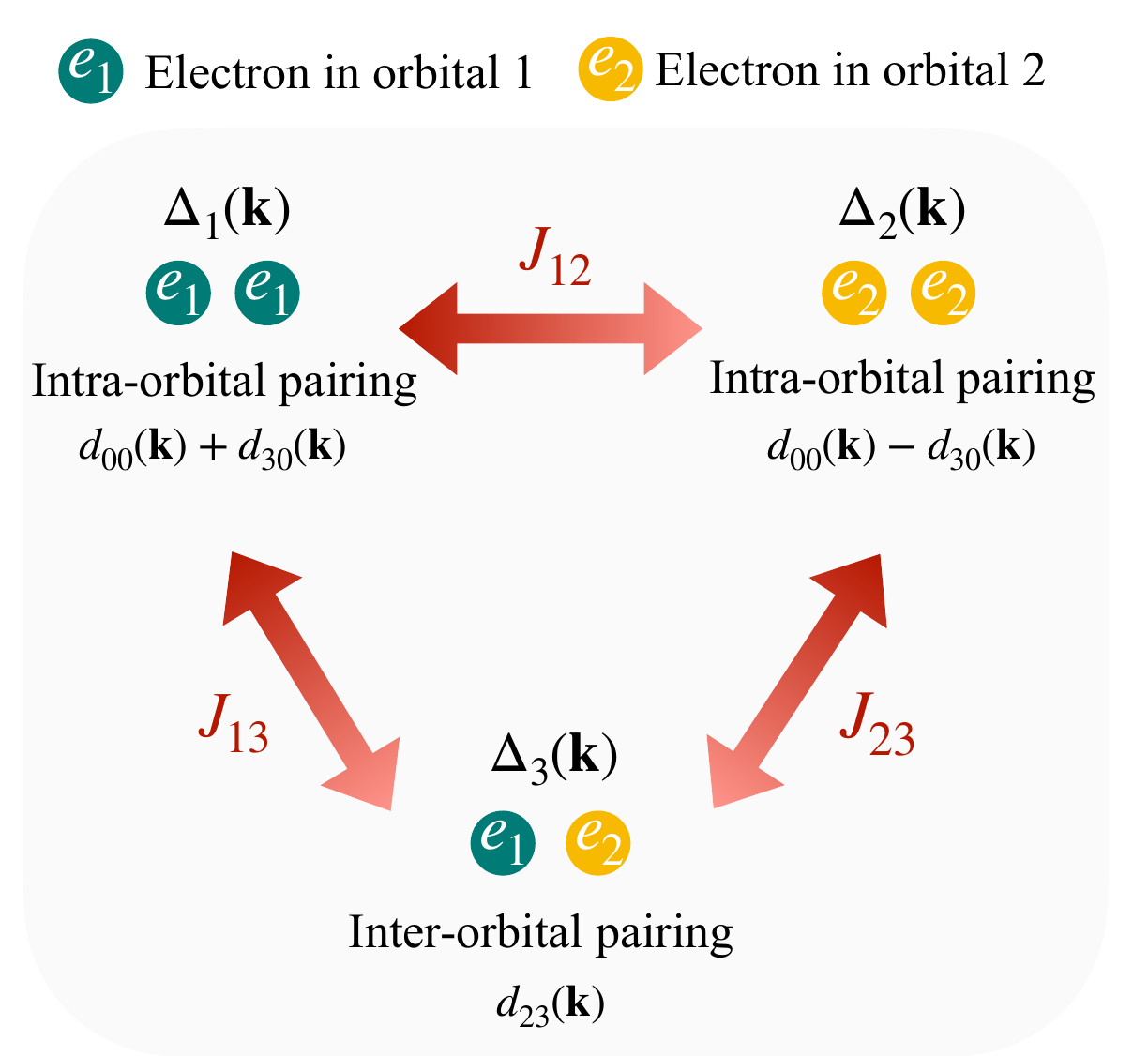}
    \caption{
    Schematic representation of the three pairing functions in a $\{d_{xz}, d_{yz}\}$-orbital system.
    $J_{12}, J_{13},$ and $J_{23}$ indicate the three Josephson couplings relating the three superconducting order parameter components $\Delta_i(\bb{k})$. Two components are related to intra-orbital spin-singlet pairing, and one is related to inter-orbital spin-triplet pairing. Indicated are also the correspondence between the two order-parameter notations used in this work, based on $d_{ab}(\bb{k})$ or $\Delta_i(\bb{k})$ parametrizations.  }
    \label{fig1}
\end{figure}

\emph{$d$-electron systems}--
Motivated by the intriguing reports of TRSB in  systems with orthorhombic symmetry~\cite{hayes_multicomponent_2021,wei_interplay_2022, hillier_nonunitary_2012, weng_twogap_2016, ghosh_quantitative_2020,hillier_evidence_2009, quintanilla_relativistic_2010}, here we discuss the case of d-electron systems with $\{d_{xz}, d_{yz}\}$ orbitals in the $D_{2h}$ point group, illustrating how, even in this simple two-orbital system, phase frustration can be tuned by changes on the Fermi surfaces.

For a two-orbital system of equal parity with $D_{2h}$ symmetry, the normal-state Hamiltonian allows intra-orbital hopping terms $h_{00}(\mathbf{k})$ and $h_{30}(\mathbf{k})$, inter-orbital hopping $h_{10}(\mathbf{k})$, and spin-orbit couplings $h_{2i}(\mathbf{k})$, with $i=\{1,2,3\}$.
Here, we focus on superconducting order parameters associated with the trivial, $A_{1g}$, irreducible representation of the point group $D_{2h}$ \cite{ramires_symmetry_2022}. A similar treatment could be extended to other symmetry channels. For simplicity, we focus on the $\bb{k}$-independent components, namely $d_{00}(\bb{k}), d_{23}(\bb{k}),$ and $d_{30}(\bb{k})$, which we treat as constants. 
To get a better physical intuition, we will now discuss the emergence of phase frustration by rewriting the superconducting order parameters in each orbital component: $\Delta_{1} = {d_{00}+ d_{30}}$, and $\Delta_{2} = {d_{00}- d_{30}}$, while the interorbital component is simply $\Delta_{3} = d_{23}$. See Fig.\ref{fig1} for a schematic illustration of this relabeling. 
The corresponding Josephson terms, that is, $J_{\alpha, \beta} \left( \Delta_\alpha(\bb{k}) \Delta^*_\beta(\bb{k}) + c.c. \right)$, can be obtained from Eqs. \eqref{2ndorderGL_T} and \eqref{t1}-\eqref{t4}. Dropping the subleading terms in $\mathcal{L}_2(T)$, they read
\begin{eqnarray}
    J_{13}&&= \frac{\mathcal{L}_1(T)}{2} \sum_{i=1,2} (-)^{i} N_i \langle \tilde{h}_{23}(\bb{k})[1 + \tilde{h}_{30}(\bb{k}) ]\rangle_{FS_i}; \nonumber \\
    J_{23}&&= \frac{\mathcal{L}_1(T)}{2} \sum_{i=1,2} (-)^{i} N_i \langle \tilde{h}_{23}(\bb{k}) [1  - \tilde{h}_{30}(\bb{k}) ]\rangle_{FS_i}; \nonumber \\
    J_{12}&&= - \frac{\mathcal{L}_1(T)}{2} \sum_{i=1,2} N_i \langle 1 -  \tilde{h}^2_{30}(\bb{k}) \rangle_{FS_i}. 
    \label{Josephson_orbital}
\end{eqnarray}

Assuming the Josephson couplings arising from the pairing interaction terms --second line in Eq.\eqref{f_GL}-- are not frustrated, when the Josephson terms listed above become dominant, a necessary condition for phase frustration requires
$J_{12} J_{13} J_{23} >0$. Since, by construction, $\langle 1 -  \tilde{h}^2_{30}(\bb{k}) \rangle_{FS_i} \geq 0$, it follows that $J_{12}\leq 0$. Consequently, the condition to have phase frustration reduces to $J_{13} J_{23} <0$.
By inspecting Eq.~\eqref{Josephson_orbital}, two minimal requirements for phase frustration can be immediately identified:  
(i) a finite spin–orbit coupling, $\tilde{h}_{23}(\mathbf{k}) \neq 0$, and  
(ii) an imbalance in the intra-orbital hopping between the two orbitals, i.e., $\tilde{h}_{30}(\mathbf{k}) \neq 0$.  
In fact, if $\tilde{h}_{23}(\mathbf{k}) = 0$, then $J_{13}=J_{23}=0$, while if $\tilde{h}_{30}(\mathbf{k}) = 0$, one finds $J_{13}=J_{23}$, such that $J_{13}J_{13}>0$. 
Given that the above requirements are satisfied, the condition to have $J_{13} J_{23} <0$ reads
\beq
\begin{split}
\left( N_{1} A_1 - N_{2}A_2\right)^2 - \left(N_{1}B_1 + N_{2}B_2\right)^2 <  0,
\end{split}
\label{condition_simpler}
\eeq
where, for conciseness, we have defined $A_{1,2}= \langle \tilde{h}_{23}(\bb{k})\rangle_{FS_{1,2}}$ and $B_{1,2}=\langle \tilde{h}_{23}(\bb{k}) \tilde{h}_{30}(\bb{k})\rangle_{FS_{1,2}}$. 
This important result suggests that the optimal condition for phase frustration is more likely realized when the quantity in the first parenthesis is the smallest, generally satisfied when the density of states in the two bands is comparable, i.e., $N_1 \sim N_2$ (assuming a constant $h_{23}(\bb{k})$, see below), and when the quantity in the second parenthesis is the largest. 

To gain more physical insight, we will now discuss how this condition is realized in a two-orbital model, which has often been used as an approximate model for iron-based superconductors~\cite{fernandes_low-energy_2016, raghu_minimal_2008}. The explicit form of $h_{ab}(\bb{k})$ terms parametrizing the normal state Hamiltonian in Eq.\eqref{H_orbspin} is
\begin{eqnarray}
   h_{00}(\bb{k})=  &&-(t_1 + t_{2o}) \cos{k_x} -(t_2 + t_{1o}) \cos{k_y}- \nonumber\\ &&  2(t_3 + t_{3o})\cos{k_x}\cos{k_y} -\mu_0; \\
   \label{h00}
    h_{30}(\bb{k})= &&-(t_1 - t_{2o}) \cos{k_x} -(t_2 - t_{1o}) \cos{k_y} \nonumber\\ &&- 2(t_3 - t_{3o})\cos{k_x}\cos{k_y} -\delta\mu_0; \\
    \label{h30}
    h_{10}(\bb{k})= &&-4t_{4}\sin{k_x}\sin{k_y};\\
    \label{h10}
    h_{23}(\bb{k})= &&\lambda,
    \label{h23}
\end{eqnarray}
where the orthorhombicity of the $D_{2h}$ lattice is taken into account by allowing for distinct onsite energies $\mu_{xz, yz} = \mu_0 \pm \delta\mu_0$, with $\delta \mu = \alpha_o \mu_0/2$, and by rescaling the hoppings as $t_{io} = (1- \alpha_o) t_i$ ($i=1,2,3$), where $\alpha_o$ quantifies the degree of orthorhombic distortion. For $\alpha_o = 0$, the model reduces to the tetragonal limit. 
In units of $|t_1| = 0.33$ eV, we set the tight-binding hopping parameters to $t_2 = -1.17$, $t_3 = 0.71$, $t_4 = 0.79$, the chemical potential to $\mu_0= 1.67$, and the spin-orbit coupling to $\lambda = 0.03$, consistent with ab-initio calculations~\cite{sknepnek_anisotropy_2009}.

With these parameters, we analyzed the range of the chemical potential $\mu$ in which the two orbital model above satisfies the condition identified in Eq.~\eqref{condition_simpler}. 
Experimentally, $\mu$ can be tuned, for example, via doping or gating. 
\begin{figure}[t!]
\includegraphics[width=\linewidth]{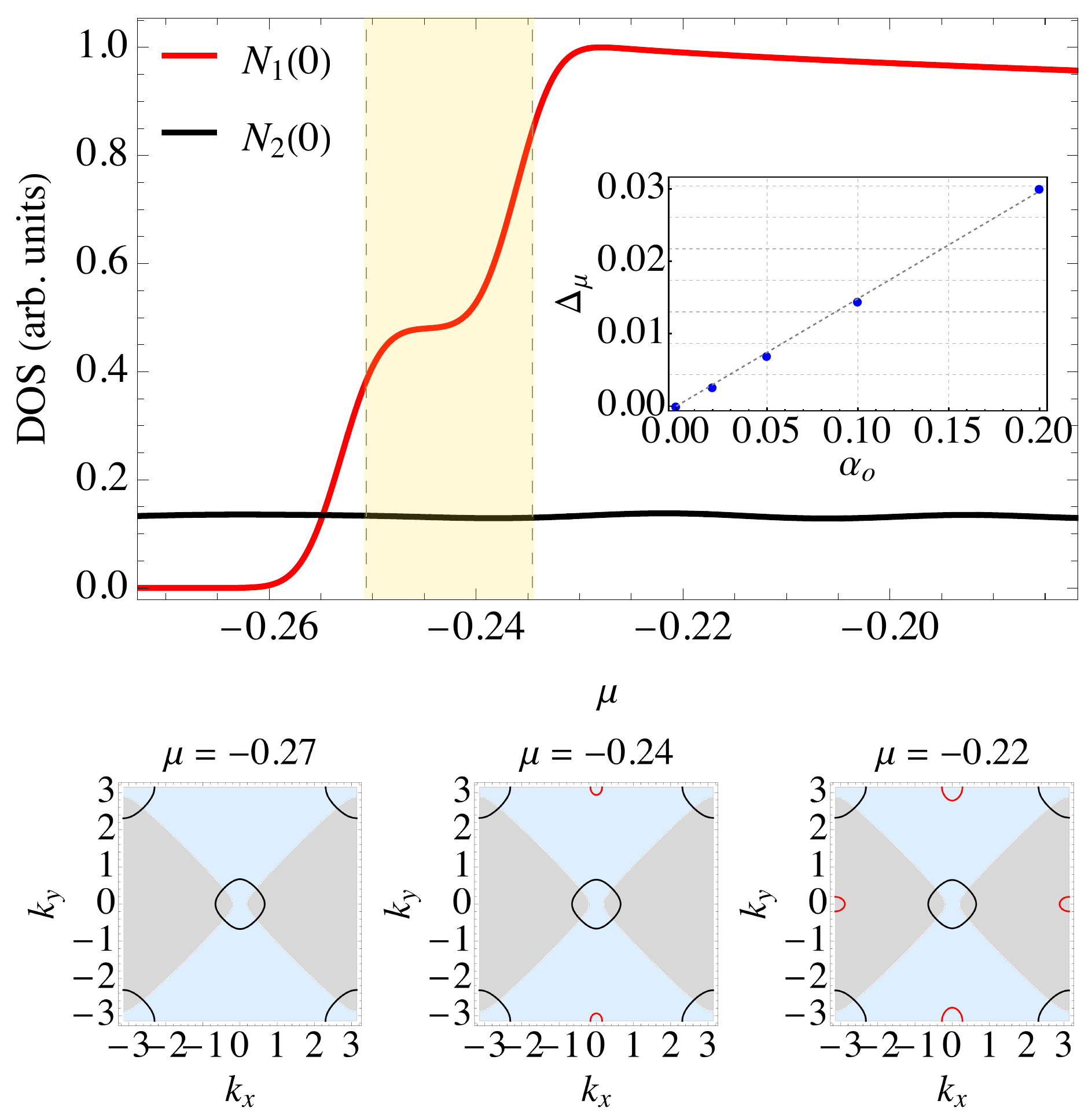}
    \caption{Upper panel: Density of states (arbitrary units) of the two bands as a function of the external chemical potential $\mu$ (in units of $|t_1| = 0.33$ eV) for the orthorhombic parameter $\alpha_o=0.1$. The parameters of the two-orbital model are given in the main text. The shaded yellow region marks the range of chemical pontential ($\Delta \mu$) within which a TRSB superconducting state emerges. Inset: $\Delta \mu$ as a function of $\alpha_o$. The dotted gray line is a linear fit, $\Delta \mu = 0.17 \alpha_o$.  Lower panels: Color map of the sign of $\tilde{h}_{23}(\mathbf{k}) \tilde{h}_{30}(\mathbf{k})$ in the $k_xk_y$ plane for three values of $\mu$, with the Fermi pockets of the two bands overlaid. {Here, gray indicates negative and light blue positive values.}}
    \label{fig:tuning_frustration_01}
\end{figure}
The results for $10\%$ orthorhombic distortion, with $\alpha_o = 0.1$, are shown in Fig.~\ref{fig:tuning_frustration_01}. 
As seen in the upper panel, the onset of a TRSB superconducting state occurs near $N_1 \sim N_2$. 
More important for the phase frustration in this scenario seems to be the maximization of the $B_{1,2}$ terms. 
In this example, as $h_{23}(\bb{k})=\lambda$ is a constant, the maximization of $B_{1,2}$ relies on the maximization of the averages of $\tilde{h}_{30}(\bb{k})$ over the respective Fermi surfaces. 
The lower panels of Fig.~\ref{fig:tuning_frustration_01} show the form factors associated with $\tilde{h}_{30}(\bb{k})$. 
While the averages over the second band (in black) do not seem to strongly depend on the chemical potential, the averages over the first band (in red) strongly depend on the chemical potential near the Lifshitz transitions.
Starting with $\mu\approx -0.22$ one sees Fermi surfaces stemming from both bands, and the averages of $h_{30}(\bb{k})$ are expected to be small given the symmetry of the form factor and the location of the Fermi surfaces in the Brillouin zone. As the chemical potential is reduced (hole doping) to $\mu\approx -0.24$, one of the Fermi pockets associated with the first band disappears, making $B_1$ significantly larger. Reducing the chemical potential further, $\mu\approx -0.27$, the remaining Fermi pocket associated with the first band disappears, such that $B_1=0$.
Our analysis suggests that orthorhombicity can be key ingredient in stabilizing TRSB superconductivity through macroscopic phase frustration {in multi-orbital d-electron systems}.

Remarkably, the condition for a TRSB superconducting state is satisfied even for very small orthorhombic distortions, as shown in the inset of Fig.~\ref{fig:tuning_frustration_01}. 
{This result highlights the robustness of the phase-frustration mechanism: even a weak breaking of the tetragonal symmetry is sufficient to lift the degeneracy between the orbital sectors and induce a finite region in parameter space within which the superconducting state breaks time-reversal symmetry. }
Such sensitivity to small lattice distortions is particularly relevant for iron-based superconductors, for which strain, nematicity, or chemical substitution naturally introduce slight anisotropies. 
Our results also provide a rationale for enlarging the parameter space within which phase frustration can occur. 
As illustrated in the inset of Fig.~\ref{fig:tuning_frustration_01}, the orthorhombic distortion acts as an effective tuning knob for the emergence of TRSB superconductivity.

Beyond the specific case of a two-orbital model with $d_{xz}$ and $d_{yz}$ orbitals within orthorhombic systems with $D_{2h}$ point group, we would like to highlight that the equations and discussion above apply to other pairs of orbitals and crystal symmetries \cite{ramires_symmetry_2022}. 
For example, for systems with $D_{2h}$ or $C_{2v}$ symmetry, if the two orbitals were $d_{xz}$ (or $d_{yz}$) and $d_{xy}$, the structure of the equations would be the same, except for the change of pairs of indices $(2,3)\rightarrow (2,1)$ [or $(2,2)$], while the very same equations would have been found for $d_{xy}$ and $d_{x^2-y^2}$ orbitals.  Interestingly, if the two orbitals are of $d_{xy}$ and $d_{x^2-y^2}$ nature, the same structure also applies to systems with $D_{2d}$ and $D_{4h}$ point groups, the latter relieving the requirement of orthorhombicity. These and other microscopic scenarios, including orbitals of opposite parity and sublattice structures are going to be explored in upcoming work.

This work establishes a microscopic foundation for realizing and controlling TRSB superconducting states in real materials. Extending this framework to include a microscopic derivation of gradient terms in the free energy will enable a deeper understanding of the topological fluctuations of the superconducting phase, offering a route to manipulate their relative nucleation energies and, ultimately, to stabilize more exotic states, such as fermion-quadrupling condensates, at finite temperature.

\begin{acknowledgments}
 \emph{Acknowledgments}--I.M. acknowledges financial support by the Swiss National Science Foundation (SNSF) via the SNSF postdoctoral Grant No. TMPFP2\_217204.
A.R. also acknowledges the SNSF, which supported the initial part of this work through Ambizione Grant No. 186043.
\end{acknowledgments}

\bibliography{MyLibrary}

\end{document}